\documentclass[fleqn,twoside]{article}
\usepackage{espcrc2}

\usepackage{graphicx}
\usepackage[figuresright]{rotating}

\newcommand{\AmS}{{\protect\the\textfont2
  A\kern-.1667em\lower.5ex\hbox{M}\kern-.125emS}}

\newcommand{\dd}{{\rm d}}
\newcommand{\MeV}{\mathrm{\ MeV}}
\newcommand{\GeV}{\mathrm{\ GeV}}

\title{A precise new KLOE measurement of $F_\pi$ with ISR events
and determination of $\pi\pi$ contribution
to $a_\mu$ for $[0.35,0.95]$ GeV$^2$}

\author{KLOE Collaboration\thanks{
F.~Ambrosino,
A.~Antonelli,
M.~Antonelli,
F.~Archilli,
C.~Bacci,
P.~Beltrame,
G.~Bencivenni,
S.~Bertolucci,
C.~Bini,
C.~Bloise,
S.~Bocchetta,
F.~Bossi,
P.~Branchini,
P.~Campana,
G.~Capon,
T.~Capussela,
F.~Ceradini,
F.~Cesario,
S.~Chi,
G.~Chiefari,
P.~Ciambrone,
F.~Crucianelli,
E.~De~Lucia,
A.~De~Santis,
P.~De~Simone,
G.~De~Zorzi,
A.~Denig,
A.~Di~Domenico,
C.~Di~Donato,
B.~Di~Micco,
A.~Doria,
M.~Dreucci,
G.~Felici,
A.~Ferrari,
M.~L.~Ferrer,
S.~Fiore,
C.~Forti,
P.~Franzini,
C.~Gatti,
P.~Gauzzi,
S.~Giovannella,
E.~Gorini,
E.~Graziani,
W.~Kluge,
V.~Kulikov,
F.~Lacava,
G.~Lanfranchi,
J.~Lee-Franzini,
D.~Leone,
M.~Martemianov,
M.~Martini,
P.~Massarotti,
W.~Mei,
S.~Meola,
S.~Miscetti,
M.~Moulson,
S.~M\"uller,
F.~Murtas,
M.~Napolitano,
F.~Nguyen,
M.~Palutan,
E.~Pasqualucci,
A.~Passeri,
V.~Patera,
F.~Perfetto,
M.~Primavera,
P.~Santangelo,
G.~Saracino,
B.~Sciascia,
A.~Sciubba,
A.~Sibidanov,
T.~Spadaro,
M.~Testa,
L.~Tortora,
P.~Valente,
G.~Venanzoni,
R.~Versaci,
G.~Xu.
}
presented by Federico Nguyen~\address{Universit\`a
degli Studi e Sezione INFN ``Roma TRE'',
Via della Vasca Navale 84, 00146 Roma, Italy}~
(\emph{e-mail address}: nguyen@fis.uniroma3.it)}
       
\begin{document}

\begin{abstract}
The KLOE experiment at the DA$\Phi$NE $\phi$-factory has
performed a new precise measurement of the pion form factor
using Initial State Radiation events, with photons emitted
at small polar angle. Results based on an integrated luminosity
of 240 pb$^{-1}$ and extraction of the
$\pi\pi$ contribution to $a_\mu$ in the mass range $[0.35,0.95]$ GeV$^2$
are presented, the systematic uncertainty is reduced with
respect to the published KLOE result.
\end{abstract}

\maketitle

\section{Introduction}
\label{sec:1}
The anomalous magnetic moment of the muon
has recently been measured to an accuracy
of 0.54 ppm~\cite{Bennett:2006fi}.
The main source of uncertainty
in the value predicted~\cite{Jegerlehner:2007xe} in the Standard Model
is given by the hadronic contribution, $a_\mu^{hlo}$,
to the lowest order.
This quantity is estimated with a dispersion
integral of the hadronic cross section measurements.

In particular, the pion form factor, $F_\pi$, defined via
$\sigma_{\pi\pi}\equiv\sigma_{e^+ e^-\to\pi^+\pi^-}\propto
s^{-1}\beta^3_\pi(s) |F_\pi(s)|^2$,
accounts for $\sim70\%$ of the central
value and for $\sim60\%$ of the uncertainty
in $a_\mu^{hlo}$.

The KLOE experiment already published~\cite{Aloisio:2004bu}
a measurement of $F_\pi$ with the method described
below, using an integrated luminosity
of 140 pb$^{-1}$, taken in 2001, henceforth
referred to as KLOE05.

\section{Measurement of $\sigma(e^+e^-\to\pi^+\pi^-\gamma)$ at DA$\Phi$NE}
\label{sec:2}
\begin{figure}
\begin{center}
\includegraphics[width=14pc,height=15.6pc]{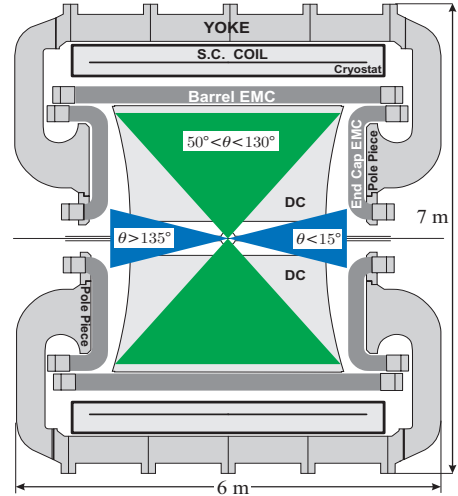}\vglue-0.5cm
\caption{Fiducial volume for the small angle photon (narrow cones)
and for the the pion tracks (wide cones).}
\label{fig:1}       
\end{center}
\end{figure}
DA$\Phi$NE is an $e^+ e^-$ collider running 
at $\sqrt{s}\simeq M_\phi$,
the $\phi$ meson mass, which has
provided 
an integrated luminosity of about 2.5 fb$^{-1}$
to the KLOE experiment up to year 2006.
In addition, about 250 pb$^{-1}$ of data have been collected
at $\sqrt{s}\simeq 1$ GeV, in 2006.
Present results are based on 240 pb$^{-1}$
of data taken in 2002. 

The KLOE detector consists of a drift chamber~\cite{Adinolfi:2002uk} with
excellent momentum resolution ($\sigma_p/p\sim 0.4\%$
for tracks with polar angle larger than $45^\circ$)
and an electromagnetic calorimeter~\cite{Adinolfi:2002zx} with good energy
($\sigma_E/E\sim 5.7\%/\sqrt{E~[\mathrm{GeV}]}$)
and precise time ($\sigma_t\sim 54~\mathrm{ps}/\sqrt{E~[\mathrm{GeV}]}\oplus
100~\mathrm{ps}$) resolution.

At DA$\Phi$NE, we measure the differential spectrum 
of the $\pi^+\pi^-$ invariant mass, $M_{\pi\pi}$, from
Initial State Radiation (ISR) events,
$e^+ e^-\to\pi^+\pi^-\gamma$, and extract
the total cross section $\sigma_{\pi\pi}\equiv\sigma_{e^+ e^-\to\pi^+\pi^-}$
using the following formula~\cite{Binner:1999bt}:
\begin{equation}
M_{\pi\pi}^2~ \frac{\dd\sigma_{\pi\pi\gamma}}
{\dd M_{\pi\pi}^2} = \sigma_{\pi\pi}
(M_{\pi\pi}^2)~ H(M_{\pi\pi}^2)~,
\label{eq:1}
\end{equation}
where $H$ is the radiator function.
This formula neglects
Final State Radiation (FSR) terms.

The cross section for ISR photons has a
divergence in the forward angle (relative to the beam
direction), such that it dominates over
FSR photon production.
The fiducial volume -- shown in Fig.~\ref{fig:1} --
is based on the following criteria:
\begin{itemize}
\item[a)] two tracks with opposite charge within the
polar angle range $50^\circ<\theta<130^\circ$; 
\item[b)] small angle photon, $\theta_\gamma<15^\circ\,
(\theta_\gamma>165^\circ)$,
the photon is not explicitly detected and its direction
is reconstructed from the track momenta in the $e^+e^-$
center of mass system,
$\vec{p}_\gamma=-(\vec{p}_{\pi^+} +\vec{p}_{\pi^-})$.
\end{itemize}
The above criteria result in events with good reconstructed
tracks and enhance the probability of having an ISR photon.
Furthermore,
\begin{itemize}
\item FSR at the Leading Order is reduced to the $0.3\%$ level;
\item the contamination from the resonant process $e^+e^-\to
  \phi\to\pi^+\pi^-\pi^0$ -- where at least one of photons coming
from the $\pi^0$ is lost -- is reduced to the level of $\sim5\%$.
\end{itemize}
Discrimination of $\pi^+\pi^-\gamma$
\begin{figure}
\begin{center}
\includegraphics[width=16.5pc, height=15.5pc]{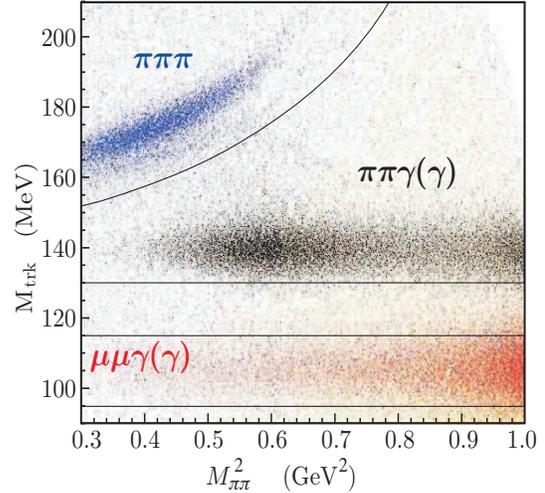}\vglue-0.5cm
\caption{Signal and background distributions in the $m_{trk}$-$M^2_{\pi\pi}$
plane; the selected area is shown.}
\label{fig:2}       
\end{center}
\end{figure}
from $e^+ e^-\to e^+ e^-\gamma$ events
is done 
via particle identification~\cite{knote:192}
based on the time of flight,
on the shape and the energy
of the clusters associated to the
tracks. In particular, electrons deposit
most of their energy in the
first planes of the calorimeter while
minimum ionizing muons and pions release uniformly the
same energy in each plane.
An event is selected if at least one
of the two tracks has not being identified
as an electron. 

Fig.~\ref{fig:2} shows that contaminations from the processes
$e^+e^-\to\mu^+\mu^-\gamma$ and
$\phi\to\pi^+\pi^-\pi^0$ are rejected
by cuts on the track mass variable, $m_{trk}$,
defined by the four-momentum conservation,
assuming a final state consisting
of two particles with the same mass and one photon

\section{Improvements with respect to the published
analysis}
\label{sec:3}
The analysis of data taken since 2002
benefits from cleaner and more stable running conditions
of DA$\Phi$NE, resulting in less machine background
and improved event filters than KLOE05.
In particular, the following changes are implemented:
\begin{itemize}
\item a new trigger level was added at the end of 2001
to eliminate the 30\% loss from pions penetrating 
to the outer calorimeter plane and thus were misidentified
as cosmic rays events. For the 2002 data, this inefficiency
has decreased down to 0.2\%, as evaluated
from a control sample;
\item the offline background filter, which 
contributed the largest experimental systematic uncertainty to the published
work~\cite{Aloisio:2004bu}, has been improved. 
The filter efficiency increased from 95\% to 98.5\%,
with negligible systematic uncertainty;
\item the vertex requirement on the two tracks
-- used in KLOE05 --
is not applied, therefore eliminating the systematic
uncertainty from this source.
\end{itemize}
The absolute normalization of the data sample is measured using
large angle Bhabha scattering events, $55^\circ<\theta<125^\circ$.
\begin{figure}[htbp]
\begin{center}
\includegraphics[width=17.5pc, height=16.5pc]{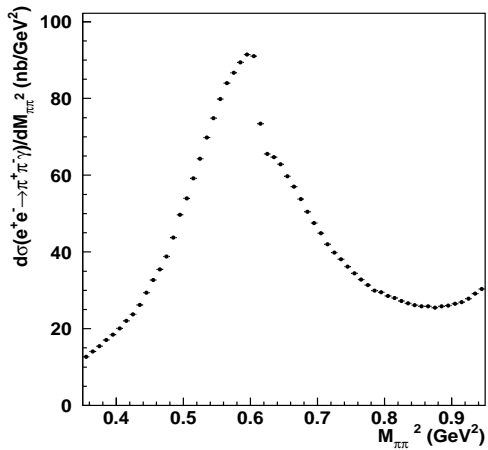}\vglue-0.5cm
\caption{Differential cross section in the $\pi\pi$ invariant mass
for the process $e^+e^-\to\pi^+\pi^-\gamma$, from an integrated
luminosity of 240 pb$^{-1}$.}
\label{fig:3}       
\end{center}
\end{figure}
The integrated luminosity, $\mathcal{L}$, is
obtained~\cite{Ambrosino:2006te} from the observed number
of events, divided by the
effective cross section evaluated from the Monte Carlo generator
\texttt{Babayaga}~\cite{Carloni Calame:2000pz},
including QED radiative corrections
with the parton shower algorithm, inserted in the code simulating
the KLOE detector. An updated version of the generator,
\texttt{Babayaga@NLO}~\cite{Balossini:2006wc},
decreased the predicted cross section
by 0.7\%, while the theoretical relative uncertainty
improved from 0.5\% to 0.1\%. The experimental relative
uncertainty on $\mathcal{L}$ is 0.3\%.

\section{Evaluation of $F_\pi$ and $a_\mu^{\pi\pi}$}
\label{sec:4}
The $\pi\pi\gamma$ differential cross section is obtained from the observed spectrum,
$N_{obs}$, after subtracting
the re\-si\-dual background events, $N_{bkg}$, and correcting for
the selection efficiency, $\varepsilon_{sel}(M_{\pi\pi}^2)$,
and the luminosity:
\begin{equation}
\frac{\dd\sigma_{\pi\pi\gamma}}
{\dd M_{\pi\pi}^2} = \frac{N_{obs}-N_{bkg}}
{\Delta M_{\pi\pi}^2}\, \frac{1}{\varepsilon_{sel}(M_{\pi\pi}^2)~ \mathcal{L}}~ ,
\label{eq:2}
\end{equation}
Fig.~\ref{fig:3} shows the
differential cross section from the selected events.

After unfolding, with the inversion
of the resolution matrix obtained from Monte Carlo,
\begin{table}
 \begin{center}
{\small
 \renewcommand{\arraystretch}{1.5}
 \setlength{\tabcolsep}{1.2mm}
\begin{tabular}{|l|c|c|}
\hline
\multicolumn{3}{|c|}{relative systematic errors on $a_\mu^{\pi\pi}$ (\%)}\\
\hline
~ & KLOE05 & KLOE08\\
\hline
offline filter & 0.6 & negligible\\
background & 0.3 & 0.6\\
$m_{trk}$ cuts & 0.2 & 0.2\\
$\pi$/e ID & 0.1 & 0.1\\ 
vertex & 0.3 & not used\\
tracking & 0.3 & 0.3\\
trigger & 0.3 & 0.1\\
acceptance & 0.3 & 0.1\\
FSR & 0.3 & 0.3\\
luminosity & 0.6 & 0.3\\
$H$  function eq.(\ref{eq:1}) & 0.5 & 0.5\\
 VP & 0.2 & 0.1\\
\hline
total & 1.3 & 1.0 \\
\hline
\end{tabular}
}
\end{center}
\caption{\label{tab:1}Comparison of systematic errors on
the extraction of $a_\mu^{\pi\pi}$ in the mass range
[0.35,0.95] GeV$^2$ between the analyses of different data sets.}
\end{table}
for events with both an initial and a final photon,
the differential cross section is corrected
using \texttt{Phokhara}
for shifting them from $M_{\pi\pi}$
to the virtual photon mass, $M_{\gamma^*}$.
Then, it is divided by the radiator function
(\texttt{Phokhara} setting the pion form factor $F_\pi=1$)
to obtain the measured total cross section $\sigma_{\pi\pi(\gamma)}(M_{\gamma^*})$,
of eq.(\ref{eq:1}).

The pion form factor is evaluated subtracting the FSR
term, $\eta_{FSR}$~\cite{Schwinger:1989ix},  
\begin{equation}
\sigma_{\pi\pi(\gamma)}~=~ \frac{\pi}{3}~
\frac{\alpha_{em}^2\,\beta_\pi^3}{M_{\gamma^*}^2}~
|F_\pi|^2 \left(1+\eta_{FSR}\right)~.
\end{equation}
The cross section for the $a_\mu^{\pi\pi}$ dispersion integral
-- inclusive of FSR --
is obtained after removing vacuum polarization, VP,
effects~\cite{Jegerlehner:2006ju}, 
\begin{equation}
\sigma_{\pi\pi(\gamma)}^{bare}=\sigma_{\pi\pi(\gamma)}
\left[\frac{\alpha_{em}(0)}{\alpha_{em}(M_{\gamma^*})}\right]^2~.
\end{equation}
Table~\ref{tab:1} shows
the list of relative systematic uncertainties
in the evaluation of $a_\mu^{\pi\pi}$ in the mass range
[0.35,0.95] GeV$^2$, for KLOE05 and for the
analysis of this new data set, KLOE08.

\section{Results}
\label{sec:5}
\begin{table}
\begin{center}
 \renewcommand{\arraystretch}{1.6}
 \setlength{\tabcolsep}{1.2mm}
\begin{tabular}{|l|c|}
\hline
\multicolumn{2}{|c|}{$a_\mu^{\pi\pi}
(M_{\pi\pi}^2\in[0.35,0.95]\GeV^2)\times10^{10}$ -- KLOE}\\
\hline
published 05 & $388.7~\pm~0.8_{stat}~\pm~4.9_{sys}$\\
\hline
updated 05 & $384.4~\pm~0.8_{stat}~\pm~4.6_{sys}$\\
\hline
new data 08 & $389.2~\pm~0.6_{stat}~\pm~3.9_{sys}$\\
\hline
\hline
\multicolumn{2}{|c|}{$a_\mu^{\pi\pi}(M_{\pi\pi}\in[630,958]\MeV)\times10^{10}$}\\
\hline
CMD-2~\cite{Ignatov:2008} & $361.5~\pm~1.7_{stat}~\pm~2.9_{sys}$\\
\hline
SND~\cite{Ignatov:2008} & $361.0~\pm~2.0_{stat}~\pm~4.7_{sys}$\\
\hline
KLOE08  & $358.0~\pm~0.6_{stat}~\pm~3.4_{sys}$\\
\hline
\end{tabular}
\end{center}
\caption{\label{tab:2}Comparison among $a_\mu^{\pi\pi}$ values evaluated with
the small $\gamma$ angle selection.}
\end{table}
The published analysis, updated
for the new Bhabha cross section and for
a bias in the trigger correction~\cite{Ambrosino:2007vj},
is compared with KLOE08, and also with the results obtained by
the VEPP--2M experiments~\cite{Ignatov:2008},
in the mass range $M_{\pi\pi}\in[630,958]\MeV$.
Table~\ref{tab:2} shows the good agreement amongst
KLOE results, and also with the published
\begin{figure}[htbp]
\begin{center}
\includegraphics[width=18pc, height=17pc]{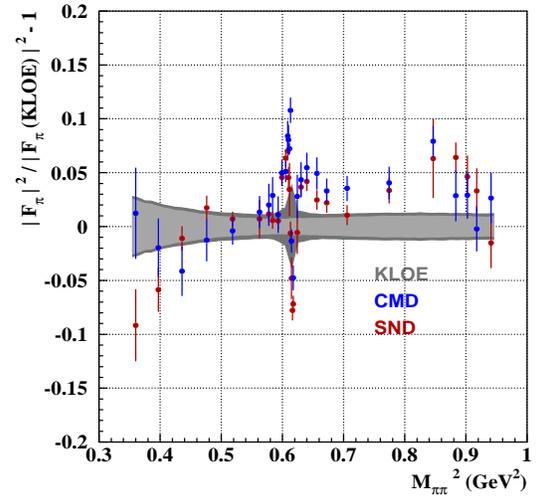}\vglue-0.5cm
\caption{Comparison of the pion form factor measured from CMD-2, SND and
KLOE, where for this latter statistical errors (light band) and
summed statistical and systematic errors (dark band) are shown.}
\label{fig:4}       
\end{center}
\end{figure}
CMD-2 and SND values. They agree with KLOE08
within one standard deviation.

The band of Fig.~\ref{fig:4} shows the KLOE08
pion form factor smoothed -- accounting for both
statistical and systematic errors -- and
normalized to fix the 0 in the ordinate scale.
CMD-2 and SND data points are interpolated
and compared to this band, in the same panel.

\section{Conclusions and outlook}
We obtained the $\pi\pi$ contribution to $a_\mu$ in the mass range
$M_{\pi\pi}^2\in[0.35,0.95]\GeV^2$ integrating
the $\pi\pi\gamma$ differential cross section for the ISR events
$e^+ e^-\to\pi^+\pi^-\gamma$, with photon emission at small angle:
\begin{enumerate}
\item KLOE08 confirms KLOE05, but with more accuracy;
\item KLOE08 is in agreement within
  one standard deviation with SND and CMD-2 values
  in the mass range $M_{\pi\pi}\in[630,958]\MeV$~\cite{Ignatov:2008}.
\end{enumerate}
Thus, $a_\mu^{\pi\pi}$ is measured to an accuracy of 0.1\%.
Independent analyses are in progress:
\begin{itemize}
\item measure $\sigma_{\pi\pi(\gamma)}$ using detected photons emitted
  at large angle, which would improve the knowledge of the FSR
  interference effects from KLOE $f_0(980)$
  measurements~\cite{Ambrosino:2005wk,Ambrosino:2006hb};
\item measure the pion form factor directly from the ratio, bin-by-bin,
  of $\pi^+\pi^-\gamma$ to $\mu^+\mu^-\gamma$ spectra~\cite{Muller:2006bk}
  (see Fig.~\ref{fig:2} for the selection of $\mu\mu\gamma$ events);
\item extract the pion form factor from
  data taken at $\sqrt{s}=1$ GeV, off the $\phi$ resonance,
  where $\pi^+\pi^-\pi^0$ background is negligible.
\end{itemize}


\begin{thebibliography}{99}
\bibitem{Bennett:2006fi}
  G.~W.~Bennett {\it et al.} [Muon G-2 Collaboration],
  Phys.\ Rev.\  D {\bf 73} (2006) 072003
\bibitem{Jegerlehner:2007xe}
  F.~Jegerlehner, ``Essentials of the Muon g-2'',
  arXiv:hep-ph/0703125
\bibitem{Aloisio:2004bu} A.~Aloisio {\it et al.} [KLOE Collaboration],
  Phys.\ Lett.\ B {\bf 606} (2005) 12
\bibitem{knote:192} A.~Denig {\it et al.} [KLOE Collaboration], 
  KLOE Note 192, July 2004,
  {\small \texttt{www.lnf.infn.it/kloe/pub/knote/kn192.ps}}
\bibitem{Adinolfi:2002uk}
  M.~Adinolfi {\it et al.}, [KLOE Collaboration]
  Nucl.\ Instrum.\ Meth.\  A {\bf 488} (2002) 51
\bibitem{Adinolfi:2002zx}
  M.~Adinolfi {\it et al.}, [KLOE Collaboration]
  Nucl.\ Instrum.\ Meth.\  A {\bf 482} (2002) 364
\bibitem{Binner:1999bt} S.~Binner, J.~H.~K\"uhn and K.~Melnikov,
  Phys.\ Lett.\  B {\bf 459} (1999) 279
\bibitem{Rodrigo:2001kf}
  G.~Rodrigo, H.~Czy\.z, J.~H.~K\"uhn and M.~Szopa,
  Eur.\ Phys.\ J.\  C {\bf 24} (2002) 71
\bibitem{Czyz:2003ue}
  H.~Czy\.z, A.~Grzelinska, J.~H.~K\"uhn and G.~Rodrigo,
  Eur.\ Phys.\ J.\  C {\bf 33} (2004) 333
\bibitem{Czyz:2006dm}
  H.~Czy\.z and E.~Nowak-Kubat,
  Phys.\ Lett.\  B {\bf 634} (2006) 493
\bibitem{Ambrosino:2006te}
  F.~Ambrosino {\it et al.} [KLOE Collaboration],
  Eur.\ Phys.\ J.\  C {\bf 47} (2006) 589
\bibitem{Carloni Calame:2000pz}
  C.~M.~Carloni Calame {\it et al.},
  Nucl.\ Phys.\  B {\bf 584} (2000) 459
\bibitem{Balossini:2006wc}
  G.~Balossini {\it et al.}, Nucl.\ Phys.\  B {\bf 758} (2006) 227
\bibitem{Jegerlehner:2006ju}
  F.~Jegerlehner, Nucl.\ Phys.\ Proc.\ Suppl.\  {\bf 162} (2006) 22
\bibitem{Schwinger:1989ix}
  J.~S.~Schwinger, ``Particles, Sources, and Fields. VOL. 3'',
{\it  Redwood City, USA: ADDISON-WESLEY (1989) 318 P.
(Advanced Book Classics Series)}
\bibitem{Ambrosino:2007vj} F.~Ambrosino {\it et al.} [KLOE Collaboration],
arXiv:0707.4078 [hep-ex]
\bibitem{Ignatov:2008} F.~Ignatov, these proceedings
\bibitem{Ambrosino:2005wk}
  F.~Ambrosino {\it et al.} [KLOE Collaboration],
  Phys.\ Lett.\  B {\bf 634} (2006) 148
\bibitem{Ambrosino:2006hb}
  F.~Ambrosino {\it et al.} [KLOE Collaboration],
  Eur.\ Phys.\ J.\  C {\bf 49} (2007) 473
\bibitem{Muller:2006bk}
  S.~E.~M\"uller and F.~Nguyen {\it et al.} [KLOE Collaboration],
  Nucl.\ Phys.\ Proc.\ Suppl.\  {\bf 162} (2006) 90
\end{thebibliography}
\end{document}